\documentclass[groupedaddress,superscriptaddress,citesort]{revtex4}
\usepackage{color}
\usepackage{bm}
\usepackage{graphicx}
\usepackage{amsbsy}
\usepackage{amsmath}
\usepackage{amsfonts}
\usepackage{amsthm}
\usepackage{amssymb}
\usepackage{tikz}
\usepackage{makecell}

\usepackage{graphicx}
\usepackage{sidecap}

\usepackage{verbatim}
\usepackage{setspace}
\usepackage{enumitem}
\setlist{nolistsep}
\begin{document}

\theoremstyle{plain}
\newtheorem{theorem}{Theorem}
\newtheorem{lemma}[theorem]{Lemma}
\newtheorem{corollary}[theorem]{Corollary}
\newtheorem{conjecture}[theorem]{Conjecture}
\newtheorem{proposition}[theorem]{Proposition}

\theoremstyle{definition}
\newtheorem{definition}{Definition}

\theoremstyle{remark}
\newtheorem*{remark}{Remark}
\newtheorem{example}{Example}

\def\be{\begin{equation}}
\def\ee{\end{equation}}
\def\ba{\begin{align}}
\def\ea{\end{align}}

\newcommand{\mE}{\mathcal{E}}
\newcommand{\mU}{\mathcal{U}}
\newcommand{\mA}{\mathcal{A}}
\newcommand{\mF}{\mathcal{F}}
\newcommand{\mI}{\mathcal{I}}
\newcommand{\mH}{\mathcal{H}}
\newcommand{\mL}{\mathcal{L}}
\newcommand{\mM}{\mathcal{M}}
\newcommand{\mT}{\mathcal{T}}
\newcommand{\mN}{\mathcal{N}}

\newcommand{\fm}{\mathcal{F}_{\bf{m}}}
\newcommand{\am}{\mathcal{A}^{\textbf{m}}}
\newcommand{\dm}{\mathcal{D}(\mathrm{H}_{{\bf m}})}
\newcommand{\lr}{\rangle\langle}
\newcommand{\la}{\langle}
\newcommand{\ra}{\rangle}
\newcommand{\tr}{{\rm Tr}}

\newcommand{\mc}[1]{\mathcal{#1}}
\newcommand{\mbf}[1]{\mathbf{#1}}
\newcommand{\mbb}[1]{\mathbb{#1}}
\newcommand{\mrm}[1]{\mathrm{#1}}

\newcommand{\bra}[1]{\langle #1|}
\newcommand{\ket}[1]{|#1\rangle}
\newcommand{\braket}[3]{\langle #1|#2|#3\rangle}
%inner product
\newcommand{\ip}[2]{\langle #1|#2\rangle}
%outer productt
\newcommand{\op}[2]{|#1\rangle \langle #2|}

\newcommand{\mbN}{\mathbb{N}}

\definecolor{eric}{rgb}{0,.5,.2}
\newcommand{\eric}[1]{{\color{eric} #1}}

\newcommand{\review}[1]{{\color{red} #1}}
\topmargin=-15mm\oddsidemargin=-2mm\textwidth=164mm\textheight=240mm
\def\ra{\rangle}
\def\la{\langle}
\def\ot{\otimes}
\def\be{\begin{equation}}
\def\ee{\end{equation}}
\def\ba{\begin{array}}
\def\ea{\end{array}}
\def\t{\tilde}
\def\Nb{{I\!\! N}}
\def\Rb{{I\!\! R}}
\def\Zb{{Z\!\!\! Z}}
\def\Fb{{I\!\! F}}
\def\Cb{{\Bbb C}}
\def\cb{{\Bbb C}}

\baselineskip=18pt

\title {Quantifying quantum non-Markovianity based on quantum coherence via skew information}
\author{Lian-He Shao}
\email{snnulhs@gmai.com}
\affiliation{School of Computer Science, Xi'an Polytechnic University, Xi'an, 710048, China}
\affiliation{State and Local Joint Engineering Research Center for Advanced Networking and Intelligent Information Services, Xi'an, 710048, China}
\author{Yu-Ran Zhang}
\email{yrzhang@csrc.ac.cn}
\affiliation{Beijing Computational Science Research Center, Beijing 100193, China}
\affiliation{Theoretical Quantum Physics Laboratory, RIKEN Cluster for Pioneering Research, Wako-shi, Saitama 351-0198, Japan}
\author{Yu Luo}
\email{luoyu@snnu.edu.cn}
\affiliation{College of Computer Science, Shaanxi Normal University, Xi'an 710062, China}
\author{Zhengjun Xi}
\email{xizhengjun@snnu.edu.cn}
\affiliation{College of Computer Science, Shaanxi Normal University, Xi'an 710062, China}
\author{Shao-Ming Fei}
\email{feishm@cnu.edu.cn}
\affiliation{School of Mathematical Sciences, Capital Normal University, Beijing 100048, China}

\begin{abstract}
Based on the nonincreasing property of quantum coherence via skew information under incoherent completely positive and trace-preserving maps, we propose a non-Markovianity measure for open quantum processes. As applications, by applying the proposed measure to some typical noisy channels, we find that it is equivalent to the three previous measures of non-Markovianity for phase damping and amplitude damping channels, i.e., the measures based on the quantum trace distance, dynamical divisibility, and quantum mutual information. For the random unitary channel, it is equivalent to the non-Markovianity measure based on $l_1$ norm of coherence for a class of output states and it is incompletely equivalent to the measure based on dynamical divisibility. We also use the modified Tsallis relative $\alpha$ entropy of coherence to detect the non-Markovianity of dynamics of quantum open systems, the results show that the modified Tsallis relative $\alpha$ entropy of coherence are more comfortable than the original Tsallis relative $\alpha$ entropy of coherence for small $\alpha$.
\end{abstract}

\maketitle

\section{Introduction}

In recent years, the Markovian and non-Markovian process for open quantum dynamics has attracted much attention. In the classical realm, the Markovian and non-Markovian dynamics are well defined and widely studied~\cite{Feller66}. However, its quantum versions are controversial in some sense~\cite{Breuer002,Rivas001,Facchi699,Zhang509}. Various non-Markovian criteria have been proposed, and some measures are introduced based on different considerations~\cite{Breuer401,Wi108,Rivas403,Chen503,Fanchini403,Kossakowski128,Lu108,Luo101,Zeng118,Zeng66,Zeng15}, such as the divisibility, mutual information, information distance measures, Fisher information flow, etc. All those non-Markovianity measures do not coincide exactly in general~\cite{song110,Addis,Haikka112,Jiang101}. Finding a universal definition for non-Markovian dynamic is an important subject in quantum information theory.

Recently, a rigorous framework of quantifying coherence has been proposed, and several measures of quantum coherence are proposed~\cite{Plenio401,Rana110,Shao120,Yu95,Rastegin136,Streltsov115,Hu2018,Streltsov89}, such as the $l_1$-norm of coherence, relative entropy of coherence, fidelity of coherence, etc.  As is well known, the Kraus operators of the phase damping channels, the amplitude channels and the random unitary channels are qubit incoherent operators, and coherence measures are monotonicity under all the incoherent completely positive and trace-preserving maps. Hence, the coherence measures are employed to investigate the detection of non-Markovianity. Chanda $et$ $al.$ used the $l_1$-norm of coherence to define the measure of non-Markovianity, and shew that it is equivalent to the measures based on trace distance, quantum mutual information and quantum divisibility for the one and two qubit phase damping channel and amplitude damping channel~\cite{Chanda1}. He $et$ $al.$ used the relative entropy of coherence to quantify the non-Markovianity and shew that the measure is also consistent with the measures based on quantum divisibility and trace distance for the qubit phase damping channel. But for the amplitude damping channel the mathematical form is still not given. To deal with this problem, they proposed an alternative non-Markovianity measure based on relative entropy of coherence in an extended space~\cite{He64,He106}.  Mirafzali $et$ $al.$ used numerical methods to detect the non-Markovianity of generalized amplitude damping model by using the Tsallis relative $\alpha$ entropy of coherence, and shew that it is consistent with the measures based on trace distance and mutual information, and is better than the non-Markovianity measure based on entanglement~\cite{Mirafzali274}.

In this paper, we use the quantum coherence via skew information to detect the non-Markovianity of quantum dynamical maps. We propose alternative non-Markovianity measures based on this coherence measure, and find that it is equivalent to the typical non-Markovianity measures based on dynamical divisibility, quantum trace distance, and quantum mutual information for the phase damping channel and the amplitude damping channel. For the random unitary channel, the proposed measure for a class of qubit states is equivalent to the measures based on the $l_1$-norm of coherence, and thus it is not completely equivalent to the measure based on dynamical divisibility. As the skew information based coherence measure is a special case of the modified Tsallis relative $\alpha$ entropy of coherence~\cite{Zhao299}, we also consider some special case for the modified Tsallis relative $\alpha$ entropy of coherence to detect the non-Markovianity of quantum dynamical maps.

The paper is organized as follows. In Sect. 2, we review basic points for quantum coherence via skew information, Tsallis relative $\alpha$ entropy of coherence and modified Tsallis relative $\alpha$ entropy of coherence, and introduce the measure of non-Markovian processes based on quantum coherence via skew information. In Sect. 3, we apply our non-Markovianity measure to some typical noisy channels. In Sect. 4. we consider some special cases for the modified Tsallis relative $\alpha$ entropy of coherence to detect the non-Markovianity both numerically and analytically. We give the summary of results in Sect. 5.

\section{quantum non-Markovianity measure based on quantum coherence via skew information}\label{II}
Firstly, we recall the basic points of quantum coherence measures. Given a finite-dimensional Hilbert space $\mathcal{H}$ with dimension $d=\dim(\mathcal{H})$, we denote $\mathcal{I}$ the set of incoherent states which are diagonal in a given basis $\{|j\rangle\}_{i=1}^d$ of $\mathcal{H}$. The incoherent quantum operations are described by the Kraus operators $\{K_n\}$ satisfying $\sum_n K_n^{\dagger}K_n= \mathbb{I}$ with $K_n \mathcal{I} K_n^\dagger\subset \mathcal{I}$. A good coherence quantifier $C(\rho)$ of a quantum state $\rho$ needs to satisfy the following conditions~\cite{Plenio401} ($i$) (Non-negative) $C(\rho)\geq 0$ for all quantum states $\rho$, and $C(\rho)=0$ if and only if $\rho\in \mathcal{I}$; ($ii$ a) (Monotonicity) for any states $\rho$ and incoherent completely positive and trace preserving (ICPTP) maps $\Phi$: $C(\rho)\geq C(\Phi(\rho)) $, where $\Phi(\rho)=\sum_n K_n \rho K_n^\dag$; ($ii$ b) (Strong monotonicity) for any states $\rho$ and incoherent operators $\{K_n\}$, $C(\rho)\geq \sum_n p_n C(\rho_n) $, where $\rho_n=\frac{K_n\rho K_n^\dag}{p_n}$ and $p_n=\mathrm{Tr}(K_n \rho K_n^\dag)$; ($iii$) (Convexity) for any ensemble $\{q_n, \rho_n\}$, $\sum_n q_n C(\rho_n)\geq C(\sum_n q_n \rho_n )$. A quantity $C(\rho)$ fulfilling the conditions ($i$)-($iii$) is called a coherence measure~\cite{Streltsov89}.

The coherence measure $C_{S}(\rho)$ via skew information is defined by~\cite{Yu95}
\begin{eqnarray}
C_{S}(\rho):=\sum_{j=1}^{d}I(\rho,|j\rangle \langle j|)=1-\sum_j \langle j|\sqrt{\rho}|j\rangle^2,
\end{eqnarray}
where $I(\rho,|j\rangle \langle j|)=-\frac{1}{2}\tr\{[\sqrt{\rho},|j\rangle \langle j|]\}^2$ corresponds to the skew information subject to the projector $|j\rangle \langle j|$.
Other two coherence quantifiers are given by the Tsallis relative $\alpha$ entropy of coherence~\cite{Rastegin136} and the modified Tsallis relative $\alpha$ entropy of coherence~\cite{Zhao299}.
The Tsallis relative $\alpha$ entropy of state $\rho$ with respect to state $\delta$ is defined by~\cite{Hiai23}
\begin{eqnarray}
T_{\alpha}(\rho\|\delta):=\frac{1}{\alpha-1}\tr(\rho^\alpha\delta^{1-\alpha}-1)
\end{eqnarray}
for $\alpha\in(0,2]$. Note that when $\alpha \rightarrow 1$, $T_{\alpha}(\rho\|\delta)$ reduces to the relative entropy $S(\rho\|\delta )=\tr(\rho\log\rho)-\tr(\rho\log\delta)$. The  coherence measure based on the Tsallis relative $\alpha$ entropy of $T_{\alpha}(\rho\|\delta)$, in the fixed reference basis $\{|j\rangle\}$, is given by~\cite{Rastegin136}
\begin{eqnarray}
C_{\alpha}(\rho)=\min_{\sigma\in \mathcal{I}}T_{\alpha}(\rho\|\sigma)=\frac{1}{\alpha-1}\{(\sum_j\langle j|\rho^\alpha|j\rangle^{\frac{1}{\alpha}})^{\alpha}-1\},
\end{eqnarray}
where $C_{\alpha}(\rho)$ satisfies all the criteria for a good coherence measure except for the strong monotonicity~\cite{Rastegin136}. Zhao $et. al.$ proposed an alternative coherence quantifier based on the modified Tsallis relative $\alpha$ entropy of coherence~\cite{Zhao299},
\begin{eqnarray}
\widetilde{C}_{\alpha}(\rho)=\min_{\sigma\in \mathcal{I}}\frac{1}{\alpha-1}[f^{\frac{1}{\alpha}}_{\alpha}(\rho,\delta)-1]=\frac{1}{\alpha-1}(\sum_j\langle j|\rho^\alpha|j\rangle^{\frac{1}{\alpha}}-1),
\end{eqnarray}
where $\alpha\in(0,2]$, $f_{\alpha}(\rho,\delta)=(\alpha-1)T_{\alpha}(\rho\|\delta)+1$, and $\{|j\rangle\}$ is the reference basis. For $\alpha=\frac{1}{2}$, $\widetilde{C}_{\frac{1}{2}}(\rho)$ reduces to $C_{S}(\rho)$.

A quantum dynamical map $\{\Phi_{t,0}\}$ is Markovian in the sense that it is completely positive divisible~(CP-divisible), i.e., $\Phi_{t,0}=\Phi_{t,\tau}\Phi_{\tau,0}$, for time $0\leq\tau\leq t$ and all completely positive  $\{\Phi\}$. Otherwise the quantum dynamical map is non-Markovian. The phase damping, amplitude and random unitary channels are all ICPTP maps. As a measure of quantum coherence, $C_{S}(\rho)$ satisfies the monotonicity under ICPTP maps, i.e. $ C_{S}(\Phi^{\mathrm{ ICPTP}}(\rho))\leq C_{S}(\rho)$. If a quantum system with the initial state $\rho(0)$ undergoes an incoherent dynamic evolution $\Phi_t^{\mathrm{ ICPTP}}$, the quantum state $\rho(t)$ at time $t$ is given by
\begin{eqnarray}
\rho(t):=\Phi_t^{\mathrm{ICPTP}} \rho(0).
\end{eqnarray}
It is easy to find that
\begin{eqnarray}
C_{S}(\rho(t))&=&C_{S}(\Phi_t^{\mathrm{ICPTP}}\rho(0)) \nonumber \\
&=&C_{S}(\Phi_{t,\tau}^{\mathrm{ICPTP}}\Phi_{\tau}^{\mathrm{ICPTP}}\rho(0)) \nonumber \\
&=&C_{S}(\Phi_{t,\tau}^{\mathrm{ICPTP}}\rho(\tau))\leq C_{S}(\rho(\tau)),
\end{eqnarray}
due to the monotonicity of $C_{S}(\rho)$ under $\mathrm{ICPTP}$ maps. Without loss of generality, $\rho(0)$ can not be chosen to be incoherent states,
which implies that $C_{S}(\rho(t))=C_{S}(\rho(\tau))=0$ for any time $t$. From Eq.~(6) we have that $C_{S}(\rho(t))$ is a monotonically decreasing function of $t$, implying that $\frac{\mathrm{d}}{\mathrm{dt}}C_{S}(\rho(t))\leq0$ for any Markovian dynamics.

From the above analysis, we can define a measure of Non-Markovianity as follow:
\begin{eqnarray}
\mathcal{N}_{C_{S}}(\Phi_t^{\mathrm{ICPTP}}):=\mathrm{sup}_{\rho(0)\in \mathcal{C}}\int_{(\mathrm{d}/\mathrm{dt})C_{S}(\rho(t))>0}\frac{\mathrm{d}}{\mathrm{dt}}C_{S}(\rho(t))\mathrm{dt},
\end{eqnarray}
where $\mathcal{C}$ is the set of all coherent states. The measure $\mathcal{N}_{C_{S}}$ can witness the non-Markovian feature only for incoherent dynamics. But it does not need auxiliary systems to easy the computation for some special coherence measures~\cite{Chanda1,He64,He106,Mirafzali274}.

\section{ Applications to open qubit systems}\label{IIII}
\subsection{Phase damping channel}
Consider a qubit system undergoing phase damping noises. The evolution is described by the differential equation
\begin{eqnarray}
\frac{\mathrm{d}}{\mathrm{dt}}\rho(t)=\gamma(t)(\sigma_z\rho(t)\sigma_z-\rho(t)),
\end{eqnarray}
with $\sigma_z$ is the standard pauli operator. The Kraus operators of Eq.~(8) can be written as $K_0(t)=\sqrt{1-\frac{h(t)}{2}}\mathbb{I}$ and $K_1(t)=\sqrt{\frac{h(t)}{2}}\sigma_z$, where $h(t)=1-f(t)$ with $f(t)=\mathrm{exp}[-2\int_0^t\gamma(\tau)\mathrm{d}\tau]$. It is easy to verify that $K_0(t)$ and $K_1(t)$ are incoherent operators in computational basis. Denote the initial state of the open quantum system as
\begin{equation}
\rho(0)=
\frac{1}{2}
\begin{pmatrix}
1+a & b  \\
b^* & 1-a
\end{pmatrix}
,
\end{equation}
where $*$ stands for the conjugation. Then the dynamics can be expressed as
\begin{equation}
\rho(t)=\Phi_t\rho(0)=
\frac{1}{2}
\begin{pmatrix}
1+a & bf(t)  \\
b^*f(t) & 1-a
\end{pmatrix}
.
\end{equation}
In order to obtain the expression of $C_{S}(\rho(t))$, we calculate the eigenvalues and eigenvectors of $\rho(t)$. After a direct calculation, the eigenvalues of $\rho(t)$ are
\begin{eqnarray}
\lambda_1=\frac{1+s(t)}{2},~~\lambda_2=\frac{1-s(t)}{2},
\end{eqnarray}
where $\sqrt{|b|^2f^2(t)+a^2}=s(t)$, with the normalized eigenvectors
\begin{eqnarray}
&&|\lambda_{1}\rangle=\left[\frac{bt}{\sqrt{2s^2(t)-2as(t)}},\frac{s(t)-a}{\sqrt{2s^2(t)-2as(t)}}\right]^{T},  \nonumber \\
&&|\lambda_{2}\rangle=\left[\frac{-bt}{\sqrt{2s^2(t)+2as(t)}},\frac{\sqrt{s(t)+a}}{\sqrt{2s(t)}}\right]^{T},
\end{eqnarray}
where $T$ stands for transpose.

Substituting $\lambda_{1}$, $\lambda_{2}$ and their normalized eigenvectors into Eq.~(1), we get
\begin{eqnarray}
C_{S}(\rho(t))=1-\Big(\sqrt{\frac{1+s(t)}{2}}\frac{s(t)+a}{2s(t)}
+\sqrt{\frac{1-s(t)}{2}}\frac{s(t)-a}{2s(t)}\Big)^2+\Big(\sqrt{\frac{1+s(t)}{2}}\frac{s(t)-a}{2s(t)}
+\sqrt{\frac{1-s(t)}{2}}\frac{s(t)+a}{2s(t)}\Big)^2.
\end{eqnarray}
After some algebraic calculation, the derivative of $C_{S}(\rho(t))$ with respect to $t$ is given by
\begin{eqnarray}
\frac{\mathrm{d}}{\mathrm{dt}}C_{S}(\rho(t))=-\frac{\gamma(t)|b|^2f^2(t)[s^4(t)-(\sqrt{1-s^2(t)}-1)^2a^2]}{s^4(t)\sqrt{1-s^2(t)}}.
\end{eqnarray}
Set $g(t)=s^4(t)-(\sqrt{1-s^2(t)}-1)^2a^2$.
Since the initial state $\rho(0)$ is coherent, the parameter $a^2<1$. Thus
$g(t)=s^4(t)\frac{(\sqrt{1-s^2(t)}+1)^2-a^2}{(\sqrt{1-s^2(t)}+1)^2}>0$.
Then $\frac{\mathrm{d}}{\mathrm{dt}}C_{S}(\rho(t))>0$ is equivalent to $\gamma(t)<0$. Namely,
the condition $\gamma(t)<0$ is a witness for non-Markovian process. The non-Markovianity measure $\mathcal{N}_{C_{S}}(\Phi_t^{\mathrm{ICPTP}})$ can be expressed as:
\begin{eqnarray}
\mathcal{N}_{C_{S}}(\Phi_t^{\mathrm{ICPTP}})=\mathrm{sup}_{\rho(0)\in \mathcal{C}}\int_{\gamma(t)<0}\frac{-\gamma(t)|b|^2f^2(t)g(t)}{s^4(t)\sqrt{1-s^2(t)}}\mathrm{dt}.
\end{eqnarray}

From the above discussion, the non-Markovianity measure based on quantum coherence via skew information matches with the Non-Markovianity measures based on the dynamical divisibility, quantum trace distance, and quantum mutual information for phase damping channel~\cite{Breuer002,Rivas001,Luo101,Zeng118}.

\subsection{Amplitude damping channel}
Consider a qubit state undergoing amplitude damping noise. The dynamics is described by the master equation,
\begin{eqnarray}
\frac{\mathrm{d}}{\mathrm{dt}}\rho(t)=-\frac{i}{2}s(t)[\sigma_+\sigma_-,\rho(t)]+\gamma(t)[\sigma_-\rho(t)\sigma_+-\frac{1}{2}\{\sigma_+\sigma_-,\rho(t)\}],
\end{eqnarray}
where $s(t)=-2\mathrm{Im}\frac{\dot{h}(t)}{h(t)}$, $\gamma(t)=-2\mathrm{Re}\frac{\dot{h}(t)}{h(t)}=-\frac{2}{|h(t)|}\frac{\mathrm{d}}{\mathrm{dt}}|h(t)|$, and the function $h(t)$ satisfies the following integrodifferential equation,
\begin{eqnarray}
\dot{h}(t)=-\int_0^tf(t-t_1)h(t_1)\mathrm{dt_1},
\end{eqnarray}
with the initial condition $h(0)=1$, $f(t-t_1)=\int J(\omega)\mathrm{exp}[i(\omega_0-\omega)(t-t_1)]\mathrm{d}\omega$ is related to the spectral density $J(\omega)$. The Krause operators of this dynamics are given by $K_0(t)=
\begin{pmatrix}
1 & 0 \\
0 & h(t)
\end{pmatrix}$ and $K_1(t)=
\begin{pmatrix}
0 & \sqrt{1-|h(t)|^2} \\
0 & 0
\end{pmatrix}$. It is easily verified that $K_{i=0,1}(t)$ are incoherent operators. The initial state is given by
\begin{equation}
\rho(0)=
\begin{pmatrix}
1-a & b  \\
b^* & a
\end{pmatrix}
,
\end{equation}
and the dynamics for $\rho(0)$ is described by
\begin{equation}
\rho(t)=\Phi_t\rho(0)=
\begin{pmatrix}
1-|h(t)|^2a & bh(t)  \\
b^*h^*(t) & |h(t)|^2a
\end{pmatrix}
.
\end{equation}

In order to obtain the expression of $C_{S}(\rho(t))$, we calculate the eigenvalues and eigenvectors of $\rho(t)$. Denote $\sqrt{(2a|h(t)|^2-1)^2+4|b|^2|h(t)|^2}=\widetilde{s}(t)$. The eigenvalues of $\rho(t)$ are
\begin{eqnarray}
\lambda_1=\frac{1+\widetilde{s}(t)}{2},~~\lambda_2=\frac{1-\widetilde{s}(t)}{2}.
\end{eqnarray}
Their normalized eigenvectors are
\begin{eqnarray}
|\lambda_{1}\rangle=\left[\frac{2bh(t)}{\sqrt{2\widetilde{s}(t)(\widetilde{s}(t)+2a|h(t)|^2-1)}},
\frac{\sqrt{\widetilde{s}(t)+2a|h(t)|^2-1}}{\sqrt{2\widetilde{s}(t)}}\right]^{T},
\end{eqnarray}
\begin{eqnarray}
|\lambda_{2}\rangle=\left[\frac{-2bh(t)}{\sqrt{2\widetilde{s}(t)(\widetilde{s}(t)-2a|h(t)|^2+1)}},
\frac{\sqrt{\widetilde{s}(t)-2a|h(t)|^2+1}}{\sqrt{2\widetilde{s}(t)}}\right]^{T}.
\end{eqnarray}
Substituting $\lambda_{1}$, $\lambda_{2}$ and their normalized eigenvectors into Eq.(1), we have
\begin{eqnarray}
C_{S}(\rho(t))&=&1-\Big(\sqrt{\frac{1+\widetilde{s}(t)}{2}}\frac{\widetilde{s}(t)-2a|h(t)|^2+1}{2\widetilde{s}(t)}
+\sqrt{\frac{1-\widetilde{s}(t)}{2}}\frac{\widetilde{s}(t)+2a|h(t)|^2-1}{2\widetilde{s}(t)}\Big)^2  \nonumber \\
&+&\Big(\sqrt{\frac{1+\widetilde{s}(t)}{2}}\frac{\widetilde{s}(t)+2a|h(t)|^2-1}{2\widetilde{s}(t)}
+\sqrt{\frac{1-\widetilde{s}(t)}{2}}\frac{\widetilde{s}(t)-2a|h(t)|^2+1)}{2\widetilde{s}(t)}\Big)^2.
\end{eqnarray}
After some algebraic calculation, the derivative of $C_{S}(\rho(t))$ is given by
\begin{center}
$\frac{\mathrm{d}}{\mathrm{dt}}C_{S}(\rho(t))=\frac{-4|b|^2|h(t)|^2\frac{\mathrm{d}}{\mathrm{dt}}|h(t)|[|h(t)|(a-|b|^2)(4a^2|h(t)|^4-8a|h(t)|^2+3)-4|b|^4|h(t)|^3
+(4a^2|h(t)|^4-1)\sqrt{a-a^2|h(t)|^2-|b|^2|}]}{\sqrt{a-a^2|h(t)|^2-|b|^2}(4a^2|h(t)|^4-4a|h(t)|^2+4|b|^2|h(t)|^2+1)^2}.$
\end{center}

Set $\widetilde{g}(t)=\frac{[|h(t)|(a-|b|^2)(4a^2|h(t)|^4-8a|h(t)|^2+3)-4|b|^4|h(t)|^3+(4a^2|h(t)|^4-1)\sqrt{a-a^2|h(t)|^2-|b|^2|}]}{\sqrt{a-a^2|h(t)|^2-|b|^2}(4a^2|h(t)|^4-4a|h(t)|^2+4|b|^2|h(t)|^2+1)^2}$. One can prove that  $\widetilde{g}(t)<0$, see Appendix A. Then the non-Markovianity measure $\mathcal{N}_{C_{S}}(\Phi_t^{\mathrm{ICPTP}})$ is given by
\begin{eqnarray}
\mathcal{N}_{C_{S}}(\Phi_t^{\mathrm{ICPTP}})=-\mathrm{sup}_{\rho(0)\in\mathcal{C}}\int_{\frac{\mathrm{d}}{\mathrm{dt}}|h(t)|>0}4|b|^2|h(t)|^2\widetilde{g}(t)\frac{\mathrm{d}}{\mathrm{dt}}|h(t)|\mathrm{dt}.
\end{eqnarray}

The condition $\mathcal{N}_{C_{S}}(\Phi_t^{\mathrm{ICPTP}})>0$ is equivalent to $\frac{\mathrm{d}}{\mathrm{dt}}|h(t)|>0$, which also matches with the Non-Markovianity measures based on the dynamical divisibility, quantum trace distance, and quantum mutual information for amplitude damping channel~\cite{Breuer002,Rivas001,Luo101,Zeng118}.

\subsection{Random unitary channel}
Now consider a qubit state undergoing random unitary noise. The dynamics is described by the master equation,
\begin{eqnarray}
\frac{\mathrm{d}}{\mathrm{dt}}\rho(t)=\sum_{i=1}^3\gamma_i(t)(\sigma_i\rho(t)\sigma_i-\rho(t)),
\end{eqnarray}
where $\gamma_i(t)$ are suitable real functions, and $\sigma_i$ are the standard three Pauli matrices. Denote the initial state of the open quantum system as
\begin{equation}
\rho(0)=%\frac{1}{2}(\mathbb{I}+\overrightarrow{r}\overrightarrow{\sigma})=
\frac{1}{2}
\begin{pmatrix}
1+r_3 & r_1-ir_2  \\
r_1+ir_2 & 1-r_3
\end{pmatrix}.
\end{equation}
The evolution of the state is give by
\begin{equation}
\rho(t)=
\frac{1}{2}\mathbb{I}+\frac{1}{2}e^{-2(\Gamma_1(t)+\Gamma_2(t))}
\begin{pmatrix}
r_3 & \omega(t)  \\
\omega^*(t) & -r_3
\end{pmatrix}
,
\end{equation}
where $\omega(t)=e^{-2\Gamma_3(t)}(e^{2\Gamma_1(t)}r_1-ie^{2\Gamma_2(t)}r_2)$.
It is generally difficult to give the mathematical form of the non-Markovianity of random unitary channel for arbitrary initial states by using the quantum coherence via skew information.
\begin{proposition}
For an initial qubit state $\rho(0)$ undergoing some qubit channels, if the output state $\rho(t)$ has the form
$\rho(t)=\begin{pmatrix}
\frac{1}{2} & bf(t)  \\
b^*f^*(t)& \frac{1}{2}
\end{pmatrix}$,
then $\frac{\mathrm{d}}{\mathrm{dt}}C_S(\rho(t))>0$ is equivalent to $\frac{\mathrm{d}}{\mathrm{dt}}C_{l_1}(\rho(t))>0$,
where $C_{l_1}(\rho)=2|b||f(t)|$ is $l_1$ norm of coherence~\cite{Plenio401}.
\end{proposition}

According to the definition of quantum coherence via skew information, we have that $C_S(\rho(t))=\frac{1}{2}-\frac{\sqrt{1-|b|^2|f(t)|^2}}{4}=\frac{1}{2}-\frac{\sqrt{4-C^2_{l_1}(\rho(t))}}{8}$ and $\frac{\mathrm{d}}{\mathrm{dt}}C_S(\rho(t))=\frac{C_{l_1}(\rho(t))}{8\sqrt{4-C^2_{l_1}(\rho(t))}}\frac{\mathrm{d}}{\mathrm{dt}}C_{l_1}(\rho(t))$, thus  $\frac{\mathrm{d}}{\mathrm{dt}}C_S(\rho(t))>0$ is equivalent to $\frac{\mathrm{d}}{\mathrm{dt}}C_{l_1}(\rho(t))>0$. Since the criterion of detecting non-Markovianity using $l_1$ norm of coherence is $\gamma_1(t)+\gamma_3(t)<0$ and $\gamma_2(t)+\gamma_3(t)<0$~\cite{He64}, if the output state is the form of $\rho(t)$ given in Proposition 1, then the criteria of detecting non-Markovianity by quantum coherence via skew information is also $\gamma_1(t)+\gamma_3(t)<0$ and $\gamma_2(t)+\gamma_3(t)<0$. It is not completely equivalent to the measure based on dynamical divisibility, in which the criteria of detecting non-Markovianity is $\gamma_1(t)+\gamma_2(t)<0$, $\gamma_1(t)+\gamma_3(t)<0$ and $\gamma_2(t)+\gamma_3(t)<0$~\cite{Chru}.

\section{ quantum non-Markovianity measure based on modified Tsallis relative $\alpha$ entropy of coherence }

In this section, we study the detection of non-Markovianity based on modified Tsallis relative $\alpha$ entropy of coherence $\widetilde{C}_{\alpha}(\rho)=\frac{1}{\alpha-1}(\sum_j\langle j|\rho^\alpha|j\rangle^{\frac{1}{\alpha}}-1)$~\cite{Zhao299}. Generally it is difficult to obtain analytical results for arbitrary $\alpha$. We consider the case for $\alpha=2$. Here we use  the same dynamics maps and initial state as the ones considered in the previous section.

[Phase damping channel]
Consider the quantum dynamical maps Eq.(8) and the initial state Eq.(9). For $\alpha=2$, according to the Eq.(4), we have
\begin{eqnarray}
\widetilde{C}_{2}(\rho(t))=\frac{\sqrt{1+2a+s^2(t)}}{2}+\frac{\sqrt{1-2a+s^2(t)}}{2}-1,
\end{eqnarray}
where $s(t)=\sqrt{|b|^2f^2(t)+a^2}$. After some algebraic calculation, we get
\begin{eqnarray}
\frac{\mathrm{d}}{\mathrm{dt}}\widetilde{C}_{2}(\rho(t))=-\frac{\gamma(t)|b|^2f^2(t)[\sqrt{s^2(t)+2a+1}+\sqrt{s^2(t)-2a+1}]}{\sqrt{s^2(t)+2a+1}\sqrt{s^2(t)-2a+1}}.
\end{eqnarray}
The non-Markovianity measure $\mathcal{N}_{\widetilde{C}_{2}}(\Phi_t^{\mathrm{ICPTP}})$ is given by
\begin{eqnarray}
\mathcal{N}_{\widetilde{C}_{2}}(\Phi_t^{\mathrm{ICPTP}})=-\mathrm{sup}_{\rho(0)\in\mathcal{C}}\int_{\gamma(t)<0}\gamma(t)\frac{|b|^2f^2(t)[\sqrt{s^2(t)+2a+1}+\sqrt{s^2(t)-2a+1}]}{\sqrt{s^2(t)+2a+1}\sqrt{s^2(t)-2a+1}}\mathrm{dt}.
\end{eqnarray}
It is easy to verify that $\mathcal{N}_{\widetilde{C}_{2}}(\Phi_t^{\mathrm{ICPTP}})>0$ is equivalent to $\gamma(t)<0$.

[Amplitude damping channel]
Consider the quantum dynamical maps Eq.(16) and the initial state Eq.(18). From Eq.(4), we have
\begin{eqnarray}
\widetilde{C}_{2}(\rho(t))=\sqrt{(1-|h(t)|^2a)^2+|b|^2|h(t)|^2}+\sqrt{|b|^2|h(t)|^2+|h(t)|^4a^2}-1,
\end{eqnarray}
and
\begin{center}
$\frac{\mathrm{d}}{\mathrm{dt}}\widetilde{C}_{2}(\rho(t))=\frac{|h(t)|\frac{\mathrm{d}}{\mathrm{dt}}|h(t)|[(2a^2|h(t)|^2+|b|^2)(\sqrt{1-2a|h(t)|^2+a^2|h(t)|^4+|b|^2|h(t)|^2}+(1-2a)\sqrt{a^2|h(t)|^4+|b|^2|h(t)|^2})]}
{\sqrt{a^2|h(t)|^4+|b|^2|h(t)|^2}\sqrt{1-2a|h(t)|^2+a^2|h(t)|^4+|b|^2|h(t)|^2}}.$
\end{center}

Let $G(t)=\frac{(2a^2|h(t)|^2+|b|^2)(\sqrt{1-2a|h(t)|^2+a^2|h(t)|^4+|b|^2|h(t)|^2}+(1-2a)\sqrt{a^2|h(t)|^4+|b|^2|h(t)|^2})}
{\sqrt{a^2|h(t)|^4+|b|^2|h(t)|^2}\sqrt{1-2a|h(t)|^2+a^2|h(t)|^4+|b|^2|h(t)|^2}}$, the proof of $G(t)>0$ is provided in appendix, then $\frac{\mathrm{d}}{\mathrm{dt}}\widetilde{C}_{2}(\rho(t))>0$ is equivalent to $\frac{\mathrm{d}}{\mathrm{dt}}|h(t)|>0$, and
\begin{eqnarray}
\mathcal{N}_{\widetilde{C}_{2}}(\Phi_t^{\mathrm{ICPTP}})=\mathrm{sup}_{\rho(0)\in\mathcal{C}}\int_{\frac{\mathrm{d}}{\mathrm{dt}}|h(t)|>0}|h(t)|G(t)\frac{\mathrm{d}}{\mathrm{dt}}|h(t)|\mathrm{dt}.
\end{eqnarray}
Clearly, the non-Markovianity measure $\mathcal{N}_{\widetilde{C}_{2}}(\Phi_t^{\mathrm{ICPTP}})$ also agrees with the Non-Markovianity measures based on the dynamical divisibility, quantum trace distance, and quantum mutual information for phase damping and amplitude damping channel~\cite{Breuer002,Rivas001,Luo101,Zeng118}.

For other values of $\alpha$, we consider a qubit maximally coherent state which undergoes the following dynamics with system Hamiltonian $H=H_{S}+H_{B}+H_{I}$, where the systems' Hamiltonian $H_{S}=\omega_{0} \sigma_{z} / 2$, $H_{B}=\sum_{k} \omega_{k} b_{k}^{\dagger} b_{k}$, and interaction Hamiltonian $H_{I}=\sum_{k} \sigma_{z}\left(g_{k} b_{k}^{\dagger}+g_{k}^{*} b_{k}\right)$. Here $\omega_{0}$ denotes the transition frequency of the atom with ground state $|0\rangle$ and excited state $|1\rangle$. The initial state is given as $|\psi_0\rangle=(|0\rangle+|1\rangle)/\sqrt{2}$. Given the spectral density $J(\omega)=\lambda W^{2} /\left\{\pi\left[\left(\omega-\omega_{0}\right)^{2}+\lambda^{2}\right]\right\}$, where $W$ is the transition strength and $\lambda$ is the spectral width of the coupling, we can obtain the master equation $\dot{\rho}_{S}^{I}(t)=\gamma(t)\left[\sigma_{z} \rho_{S}^{I}(t) \sigma_{z}-\rho_{S}^{I}(t)\right]$, where the time-dependent decay rate $\gamma(t)$ is given by~\cite{Breuerb}
\begin{equation}
\gamma(t)=\left\{\begin{array}{l}{\frac{4 W^{2} \sinh \left(\frac{d t}{2}\right)}{d \cosh \left(\frac{d t}{2}\right)+\lambda \sinh \left(\frac{d t}{2}\right)}, \quad W \leq \lambda / 2} \\[3mm]
{\frac{4 W^{2} \sin \left(\frac{d t}{2}\right)}{d \cos \left(\frac{d t}{2}\right)+\lambda \sin \left(\frac{d t}{2}\right)} \quad, \quad W>\lambda / 2}\end{array}\right.
\end{equation}
with $d=\sqrt{\left|\lambda^{2}-4 W^{2}\right|}$.

\begin{figure}\label{fig1}
  \includegraphics[scale=0.6]{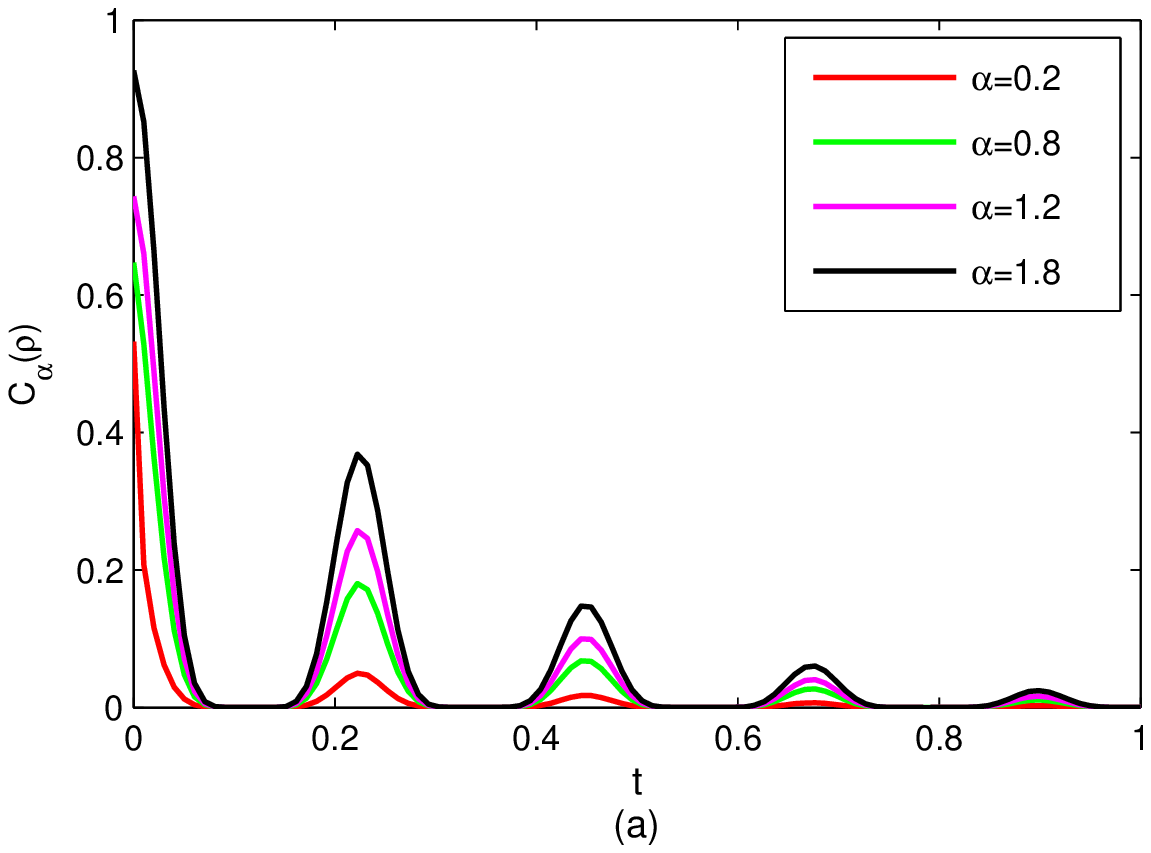}
  \includegraphics[scale=0.6]{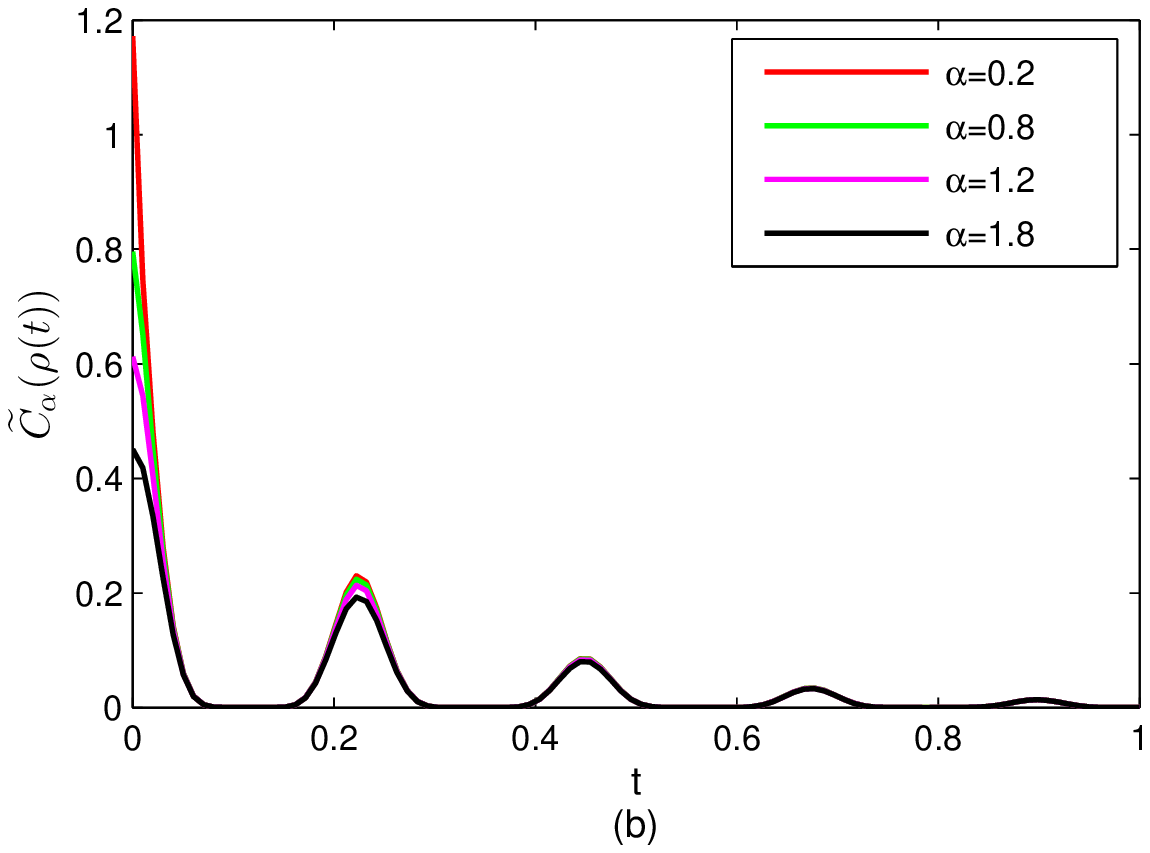}
\caption{ $C_{\alpha}(\rho(t))$ and $\widetilde{C}_{\alpha}(\rho(t))$ in terms of $t$ for different $\alpha$ ($\alpha=0.2, 0.8, 1.2, 1.8$) with $W=14$.}
\label{Fig_1}
\end{figure}

\begin{figure}\label{fig2}
  \includegraphics[scale=0.6]{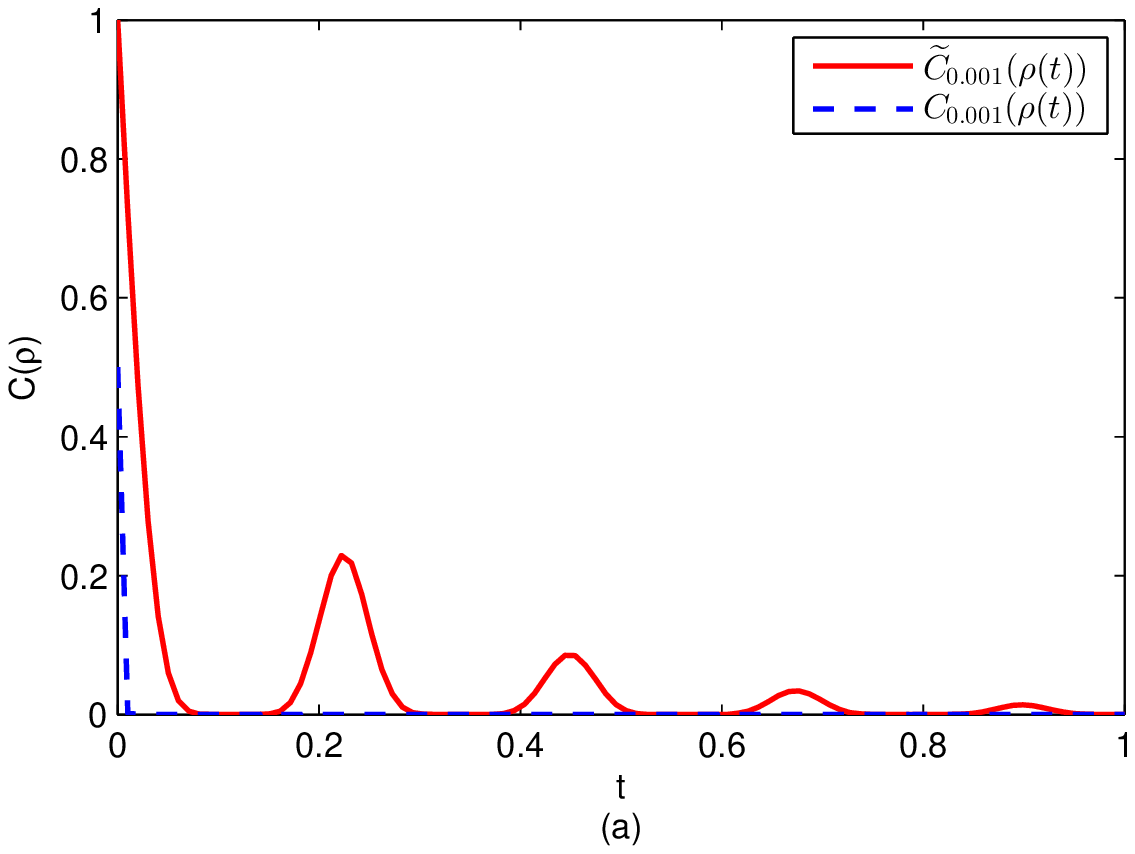}
  \includegraphics[scale=0.6]{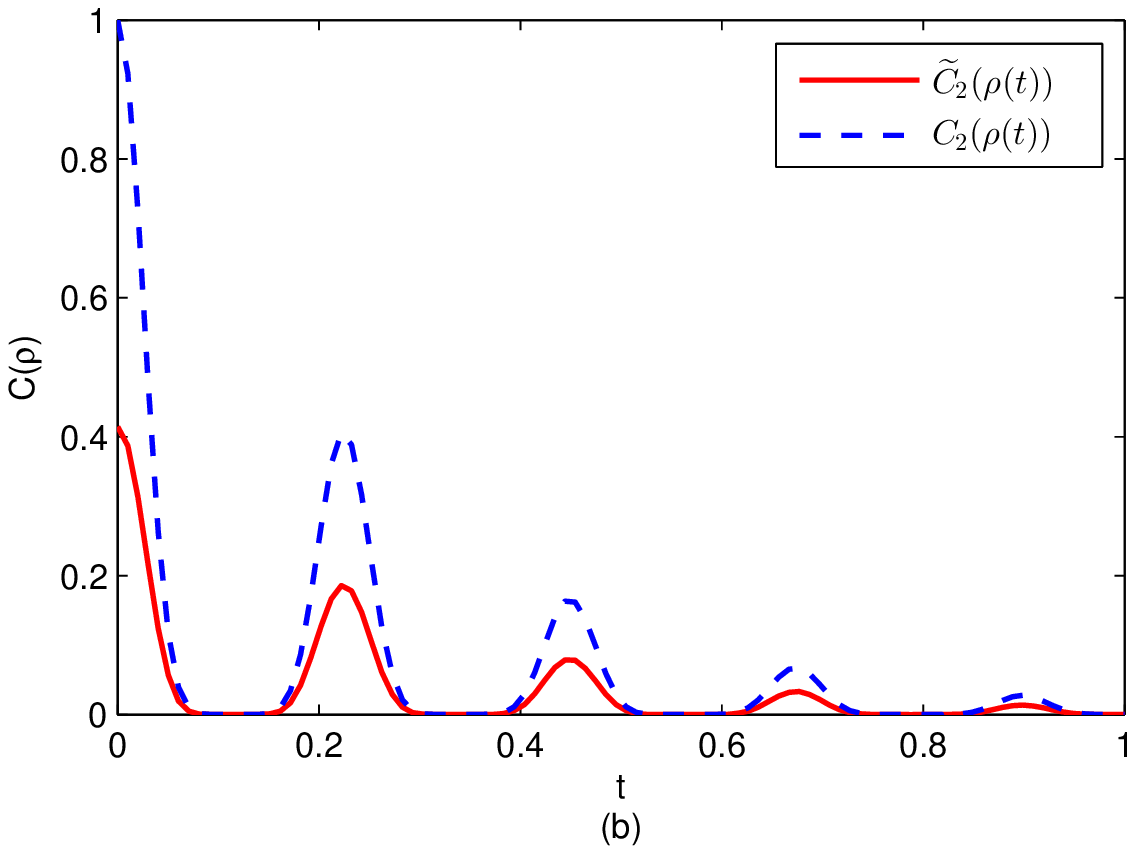}
\caption{  $C_{\alpha}(\rho(t))$ and $\widetilde{C}_{\alpha}(\rho(t))$ in terms of $t$ for small $\alpha$ ($\alpha=0.001$) and big $\alpha$ ($\alpha=2$) with $W=14$.}
\label{Fig_2}
\end{figure}

\begin{figure}\label{fig3}
  \includegraphics[scale=0.6]{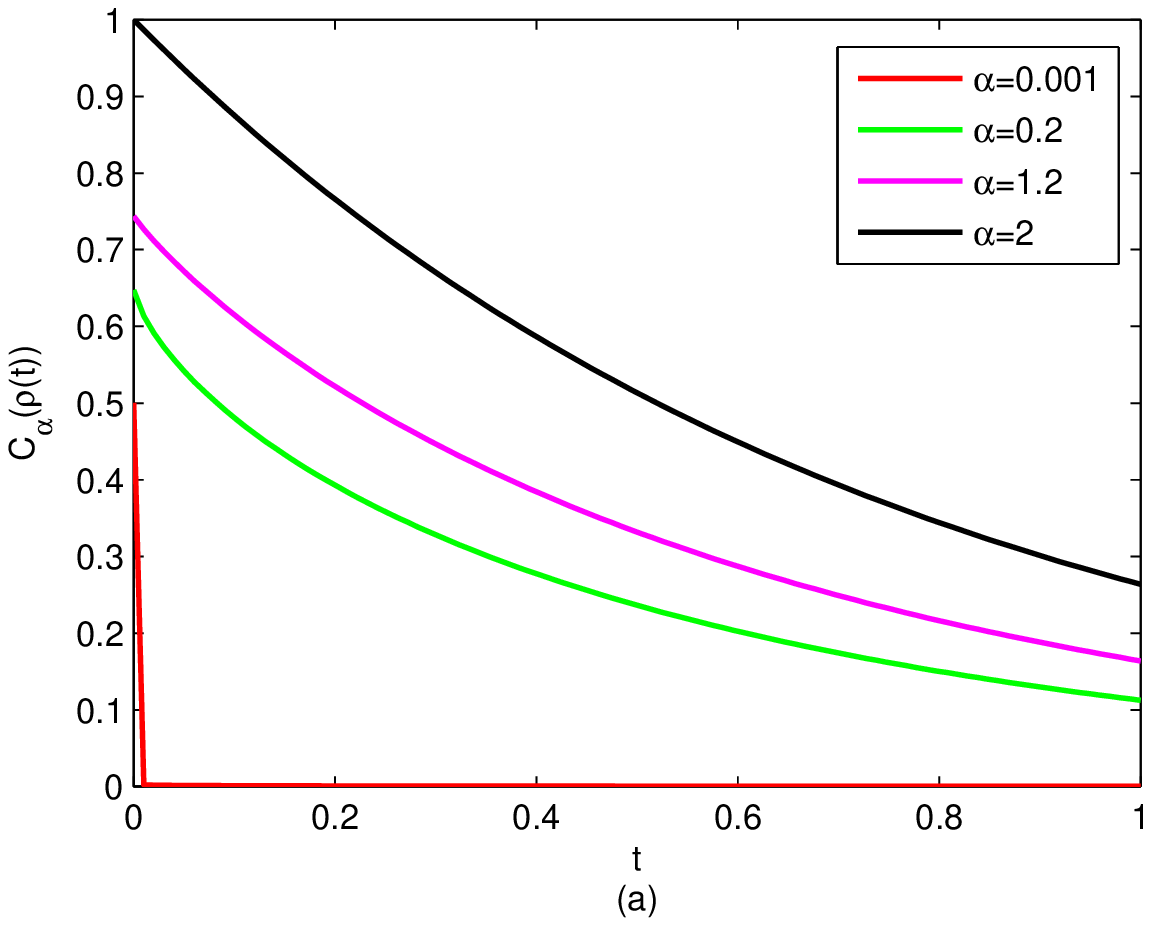}
  \includegraphics[scale=0.6]{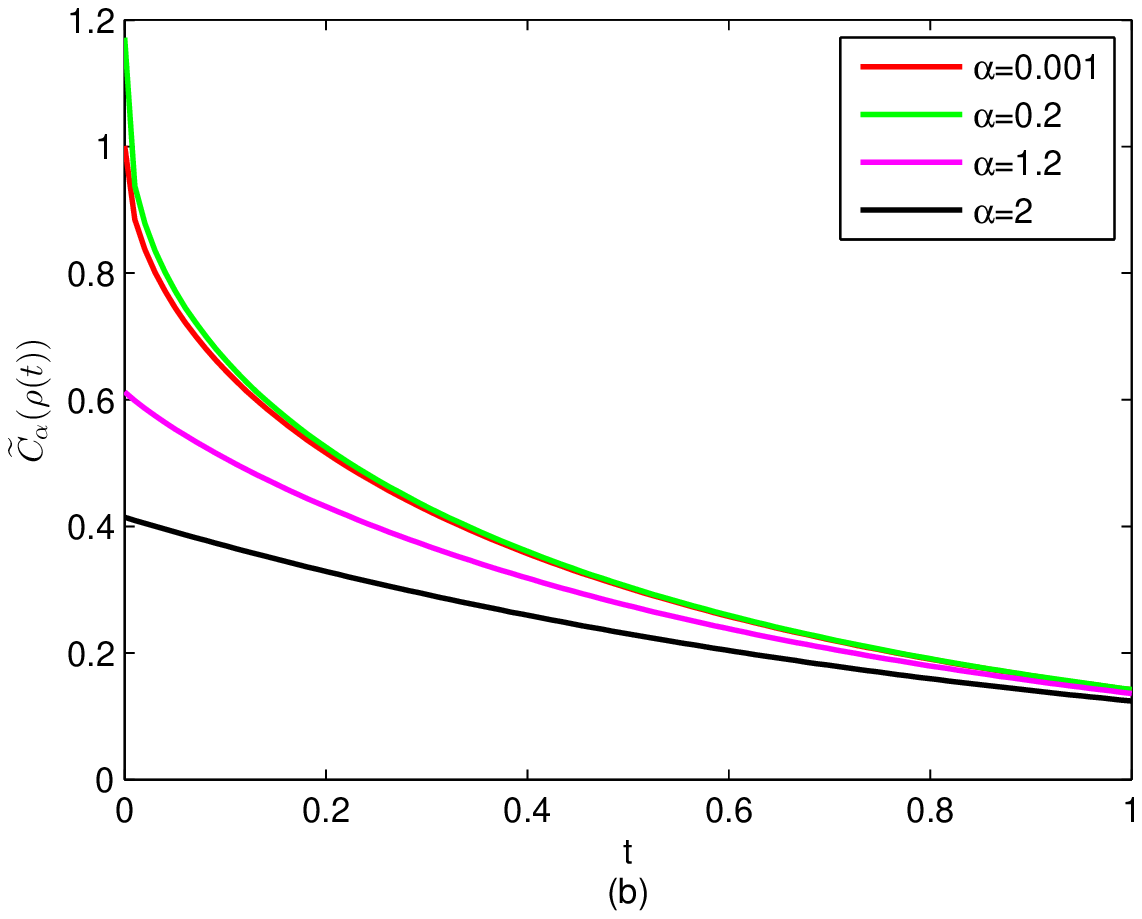}
\caption{ $C_{\alpha}(\rho(t))$ and $\widetilde{C}_{\alpha}(\rho(t))$ in terms of $t$ for different $\alpha$ ($\alpha=0.001, 0.2, 1.2, 2$) with small $W$ ($W=0.5$).}
\label{Fig_3}
\end{figure}

\begin{figure}\label{fig4}
  \includegraphics[scale=0.6]{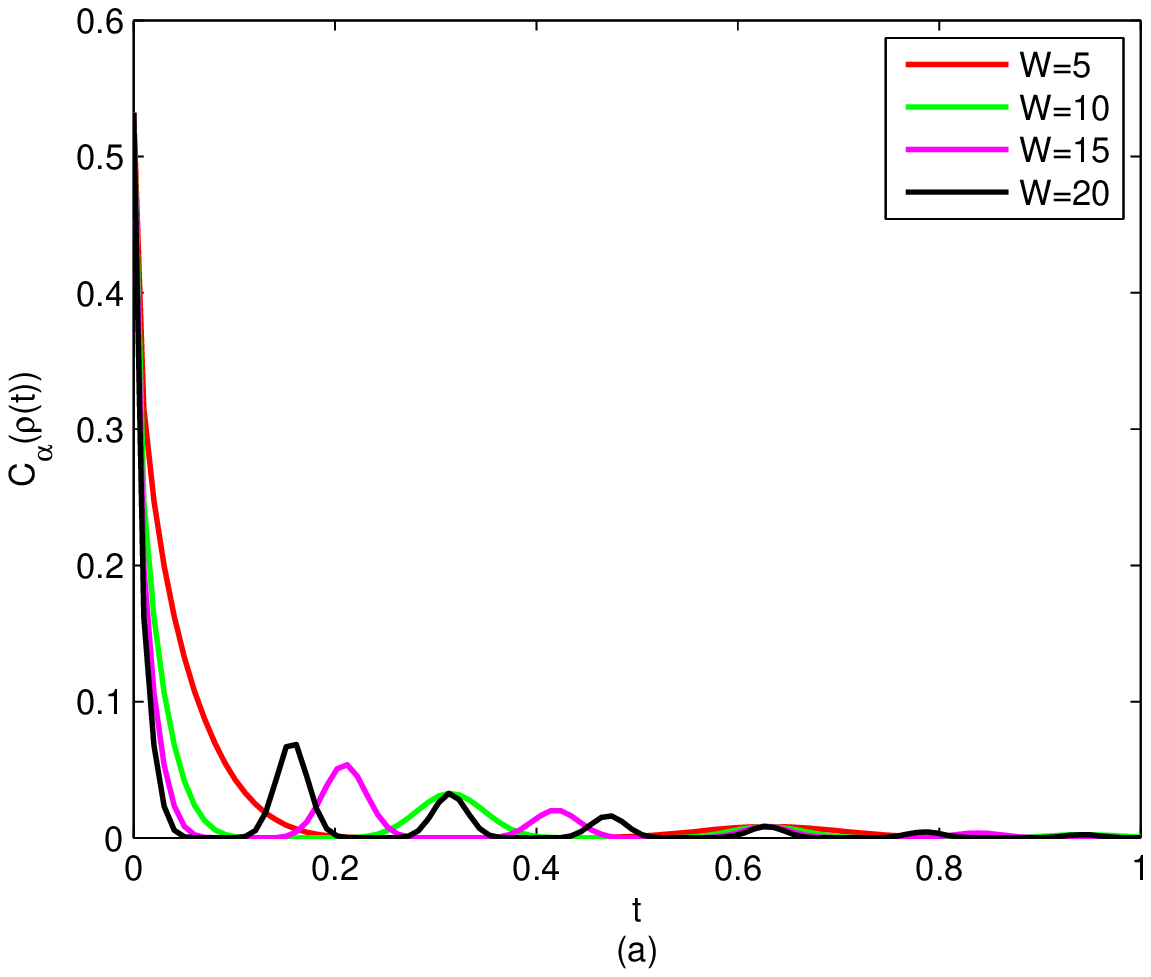}
  \includegraphics[scale=0.6]{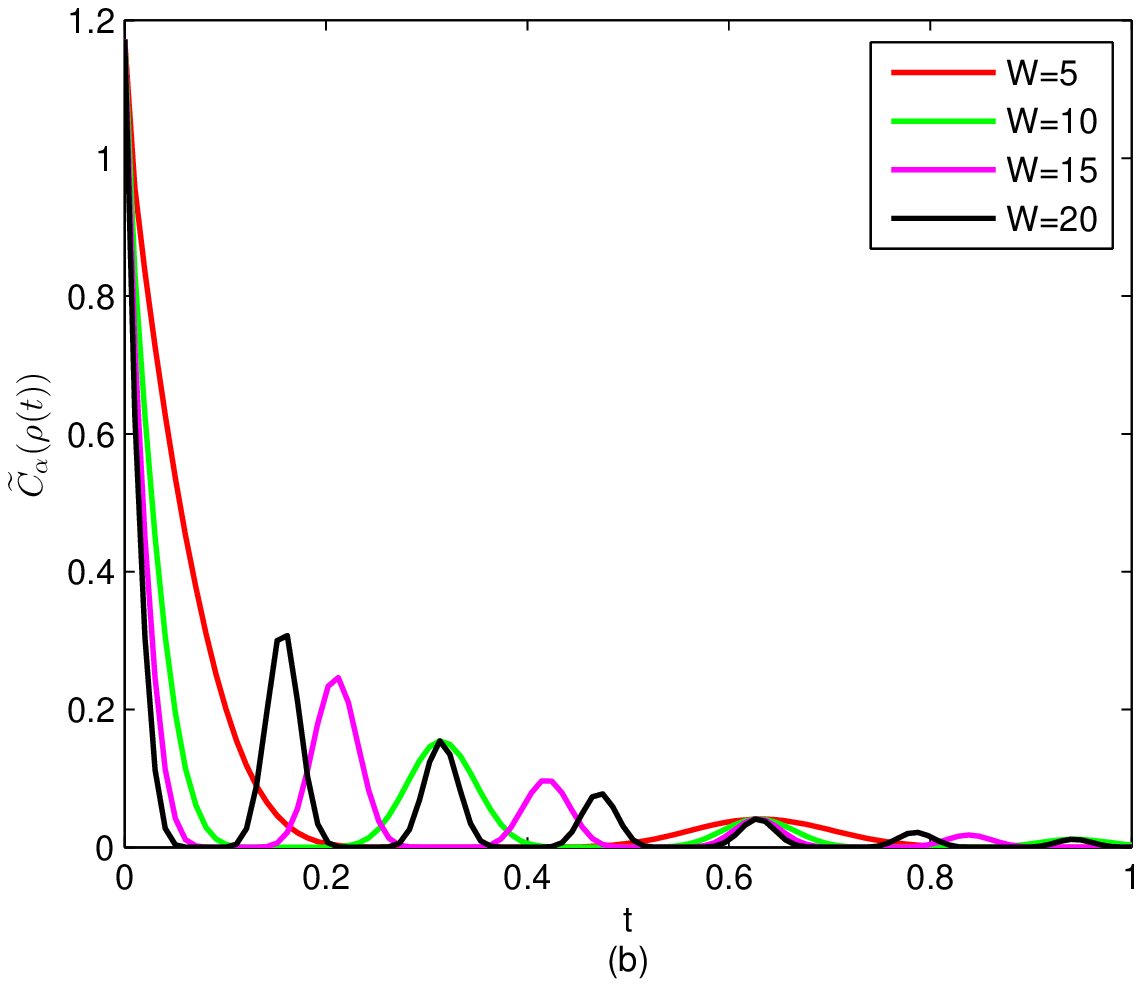}
\caption{$C_{\alpha}(\rho(t))$ and $\widetilde{C}_{\alpha}(\rho(t))$ in terms of $t$ for different $W$ ($W=5, 10, 15, 20$) with $\alpha=0.2$.}
\label{Fig_4}
\end{figure}

\begin{figure}\label{fig5}
  \includegraphics[scale=0.7]{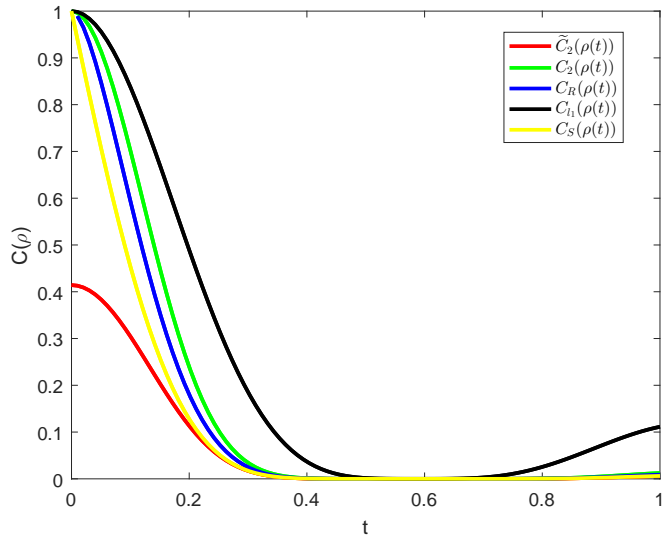}
\caption{$C_{2}(\rho(t))$, $\widetilde{C}_{2}(\rho(t))$, $C_{S}(\rho(t))$, $C_{l_1}(\rho(t))$ and $C_{R}(\rho(t))$ in terms of $t$ for $W=3$.}
\label{Fig_5}
\end{figure}

In Fig.~1, we plot $C_{\alpha}(\rho(t))$ and $\widetilde{C}_{\alpha}(\rho(t))$ in terms of $t$ for different $\alpha$ ($\alpha=0.2, 0.8, 1.2, 1.8$) with $W=14$. As shown in Fig.~1, there are many time intervals in which the monotonic decrease of $C_{\alpha}(\rho(t))$ and $\widetilde{C}_{\alpha}(\rho(t))$, have been changed to temporary increment. Therefore, $C_{\alpha}(\rho(t))$ and $\widetilde{C}_{\alpha}(\rho(t))$ can detect the non-Markovianity of this model. In Fig.~1(a), we show that for small $\alpha$, the ability to detect non-Markovianity by using $C_{\alpha}(\rho(t))$ is weaken. Fig.~1(b) shows that the ability to detect non-Markovianity by using $\widetilde{C}_{\alpha}(\rho(t))$ keeps almost the same when the value of $\alpha$ changes.

In order to explain clearly about the impact of the change of $\alpha$, we plot  $C_{\alpha}(\rho(t))$ and $\widetilde{C}_{\alpha}(\rho(t))$ in terms of $t$ for small $\alpha$ ($\alpha=0.001$) with $W=14$. In Ref.~\cite{Mirafzali274}, the authors give the result that $C_{\alpha}(\rho(t))$ is not comfortable for detection of non-Markovianity for generalized amplitude damping channel when $\alpha$ is small. In Fig.~2(a), we also show that when the value of $\alpha$ is small,  $C_{\alpha}(\rho(t))$ is not comfortable for detection of non-Markovianity for phase damping channel. As a contrast, we also plot $\widetilde{C}_{\alpha}(\rho(t))$ for small $\alpha$. One can see that using $\widetilde{C}_{\alpha}(\rho(t))$ to detect non-Markovianity is more comfortable than $C_{\alpha}(\rho(t))$. We also plot $C_{\alpha}(\rho(t))$ and $\widetilde{C}_{\alpha}(\rho(t))$ in terms of $t$ for large $\alpha$ ($\alpha=2$). Figure~2(b) shows that the large $\alpha$ is suitable for both  $C_{\alpha}(\rho(t))$ and $\widetilde{C}_{\alpha}(\rho(t))$ to detect the non-Markovianity.

In Fig.~3, we plot $C_{\alpha}(\rho(t))$ and $\widetilde{C}_{\alpha}(\rho(t))$ in terms of $t$ for different $\alpha$ ($\alpha=0.001, 0.2, 1.2, 2$) with $W=0.5$. It is shown that both $C_{\alpha}(\rho(t))$ and $\widetilde{C}_{\alpha}(\rho(t))$ do not detect the non-Markovianity in the weak coupling and Markovian regime $W\leq 0.5$.

In Fig.~4, we plot $C_{\alpha}(\rho(t))$ and $\widetilde{C}_{\alpha}(\rho(t))$ in terms of $t$ for different $W$ ($W=5, 10, 15, 20$) with $\alpha=0.2$. It is shown that, for both $C_{\alpha}(\rho(t))$ and $\widetilde{C}_{\alpha}(\rho(t))$ , with strong coupling and stronger non-Markovian regime $W$, the non-Markovianity is more obviously measured.

In Fig.~5, we plot several coherence measures for $W=3$. It is shown that, $C_{\alpha}(\rho(t))$, $\widetilde{C}_{\alpha}(\rho(t))$, $C_{S}(\rho(t))$ and relative entropy of coherence are not comfortable for detection of non-Markovianity except for the $l_1$ norm of coherence.

\section{\bf summary}\label{IIIIII}

In this paper, we have given an alternative non-Markovianity measure of incoherent quantum dynamical maps, based on quantum coherence via skew information. For phase damping and amplitude damping channels, we have shown that the non-Markovianity measure is equivalent to the three previous measures of non-Markovianity, i.e., the measures based on the quantum trace distance, dynamical divisibility, and quantum mutual information. For the random unitary channel, we also show that it is equivalent to the non-Markovianity measure based on $l_1$ norm of coherence for a class of output states and it is not completely equivalent to the measure based on dynamical divisibility. As the quantum coherence via skew information is a special case of the modified Tsallis relative $\alpha$ entropy of coherence, we then use the modified Tsallis relative $\alpha$ entropy of coherence to detect the non-Markovianity of dynamics of quantum open systems, the results have shown that the modified Tsallis relative $\alpha$ entropy of coherence can also detect non-Markovianity for phase damping and amplitude damping channels. We have numerically shown that the modified Tsallis relative $\alpha$ entropy of coherence are more comfortable than the original Tsallis relative $\alpha$ entropy of coherence for small $\alpha$ in detecting of non-Markovianity for our dynamics model.

\bigskip
\noindent {\bf Acknowledgments} This work is supported by NSFC under NOS 11847101, U1530401, 61671280 and 11675113, Xi'an Polytechnic University scientific research funds (BS201851), Young Talent fund of University Association for Science and Technology in Shaanxi, China Postdoctoral Science Foundation (Grant No.2018M640055),JSPS Postdoctoral Fellowship (grant No. P19326) and the Fundamental Research Funds for the Cenltral Universities (GK201902007),
Beijing Municipal Commission of Education (KZ201810028042) and Beijing Natural Science Foundation (Z190005).

\section{appendix}
\subsection{The proof of $\widetilde{g}(t)<0$}
\vspace{-0.1cm}
That $\frac{[|h(t)|(a-|b|^2)(4a^2|h(t)|^4-8a|h(t)|^2+3)-4|b|^4|h(t)|^3+(4a^2|h(t)|^4-1)\sqrt{a-a^2|h(t)|^2-|b|^2}]}{\sqrt{a-a^2|h(t)|^2-|b|^2}(4a^2|h(t)|^4-4a|h(t)|^2+4|b|^2|h(t)|^2+1)^2}=\widetilde{g}(t)<0$ implies that the numerator $[|h(t)|(a-|b|^2)(4a^2|h(t)|^4-8a|h(t)|^2+3)-4|b|^4|h(t)|^3+(4a^2|h(t)|^4-1)\sqrt{a-a^2|h(t)|^2-|b|^2}]<0$, which is true since
\begin{eqnarray}
 &&[|h(t)|(a-|b|^2)(4a^2|h(t)|^4-8a|h(t)|^2+3)-4|b|^4|h(t)|^3+(4a^2|h(t)|^4-1)\sqrt{a-a^2|h(t)|^2-|b|^2}]  \nonumber  \\
&&=|h(t)|(a-|b|^2)[4(a|h(t)|^2-1)^2-1]-4|b|^4|h(t)|^3+(2a|h(t)|^2+1)(2a|h(t)|^2-1)\sqrt{a-a^2|h(t)|^2-|b|^2} \nonumber \\
%&&=|h(t)|(a-|b|^2)[2(a|h(t)|^2-1)+1][2(a|h(t)|^2-1)-1]-4|b|^4|h(t)|^3 \nonumber \\
%&&+(2a|h(t)|^2+1)(2a|h(t)|^2-1)\sqrt{a-a^2|h(t)|^2-|b|^2} \nonumber \\
%&&=|h(t)|(a-|b|^2)(2a|h(t)|^2-1)(2a|h(t)|^2+1)-4|h(t)|(a-|b|^2)(2a|h(t)|^2-1)-4|b|^4|h(t)|^3 \nonumber \\
%&&+(2a|h(t)|^2+1)(2a|h(t)|^2-1)\sqrt{a-a^2|h(t)|^2-|b|^2} \nonumber \\
&&=(2a|h(t)|^2+1)(2a|h(t)|^2-1)[|h(t)|(a-|b|^2)+\sqrt{a-a^2|h(t)|^2-|b|^2}]-4|h(t)|[(a-|b|^2)(2a|h(t)|^2-1)+|b|^4|h(t)|^2] \nonumber \\
&&=(2a|h(t)|^2+1)(2a|h(t)|^2-1)[|h(t)|(a-|b|^2)+\sqrt{a-a^2|h(t)|^2-|b|^2}]-4|h(t)|[|h(t)|^2(a-|b|^2)^2-(a-a^2|h(t)|^2-|b|^2)] \nonumber \\
%&&=(2a|h(t)|^2+1)(2a|h(t)|^2-1)[|h(t)|(a-|b|^2)+\sqrt{a-a^2|h(t)|^2-|b|^2}]\nonumber \\
%&&-4|h(t)|[|h(t)|(a-|b|^2)+\sqrt{a-a^2|h(t)|^2-|b|^2}][|h(t)|(a-|b|^2)-\sqrt{a-a^2|h(t)|^2-|b|^2}] \nonumber \\
&&=[|h(t)|(a-|b|^2)+\sqrt{(a-a^2|h(t)|^2-|b|^2)}][4a^2|h(t)|^4-1-4a|h(t)|^2+4|b|^2|h(t)|^2+4|h(t)|\sqrt{a-a^2|h(t)|^2-|b|^2}] \nonumber \\
&&=[|h(t)|(a-|b|^2)+\sqrt{(a-a^2|h(t)|^2-|b|^2)}]\{-[4|h(t)|^2(a-a^2|h(t)|^2-|b|^2)-4|h(t)|\sqrt{a-a^2|h(t)|^2-|b|^2}+1] \}\nonumber \\
&&=[|h(t)|(a-|b|^2)+\sqrt{(a-a^2|h(t)|^2-|b|^2)}]\{-[4|h(t)|^2(\sqrt{a-a^2|h(t)|^2-|b|^2}-1)^2] \}<0.
\end{eqnarray}

\subsection{The proof of $G(t)>0$}
\vspace{-0.1cm}
If $\frac{[(2a^2|h(t)|^2+|b|^2)(\sqrt{a^2|h(t)|^4+|b|^2|h(t)|^2}+\sqrt{1-2a|h(t)|^2+a^2|h(t)|^4+|b|^2|h(t)|^2}-2a\sqrt{a^2|h(t)|^4+|b|^2|h(t)|^2})]}
{\sqrt{a^2|h(t)|^4+|b|^2|h(t)|^2}\sqrt{1-2a|h(t)|^2+a^2|h(t)|^4+|b|^2|h(t)|^2}}=G(t)>0$, then the numerator $[(2a^2|h(t)|^2+|b|^2)(\sqrt{a^2|h(t)|^4+|b|^2|h(t)|^2}+\sqrt{1-2a|h(t)|^2+a^2|h(t)|^4+|b|^2|h(t)|^2}-2a\sqrt{a^2|h(t)|^4+|b|^2|h(t)|^2})]>0$.

Since $(2a^2|h(t)|^2+|b|^2)(\sqrt{a^2|h(t)|^4+|b|^2|h(t)|^2}+\sqrt{1-2a|h(t)|^2+a^2|h(t)|^4+|b|^2|h(t)|^2}-2a\sqrt{a^2|h(t)|^4+|b|^2|h(t)|^2})$\\
$=\frac{(2a^2|h(t)|^2+|b|^2)^2(\sqrt{a^2|h(t)|^4+|b|^2|h(t)|^2}+\sqrt{1-2a|h(t)|^2+a^2|h(t)|^4+|b|^2|h(t)|^2})^2-4a^2(a^2|h(t)|^4+|b|^2|h(t)|^2)}
{(2a^2|h(t)|^2+|b|^2)(\sqrt{a^2|h(t)|^4+|b|^2|h(t)|^2}+\sqrt{1-2a|h(t)|^2+a^2|h(t)|^4+|b|^2|h(t)|^2})+2a\sqrt{a^2|h(t)|^4+|b|^2|h(t)|^2}}$,
%\begin{eqnarray}
%&&(2a^2|h(t)|^2+|b|^2)(\sqrt{a^2|h(t)|^4+|b|^2|h(t)|^2}+\sqrt{1-2a|h(t)|^2+a^2|h(t)|^4+|b|^2|h(t)|^2}-2a\sqrt{a^2|h(t)|^4+|b|^2|h(t)|^2})  \nonumber  \\
%&&=\frac{(2a^2|h(t)|^2+|b|^2)^2(\sqrt{a^2|h(t)|^4+|b|^2|h(t)|^2}+\sqrt{1-2a|h(t)|^2+a^2|h(t)|^4+|b|^2|h(t)|^2})^2-4a^2(a^2|h(t)|^4+|b|^2|h(t)|^2)}
%{(2a^2|h(t)|^2+|b|^2)(\sqrt{a^2|h(t)|^4+|b|^2|h(t)|^2}+\sqrt{1-2a|h(t)|^2+a^2|h(t)|^4+|b|^2|h(t)|^2})+2a\sqrt{a^2|h(t)|^4+|b|^2|h(t)|^2}} \nonumber
%\end{eqnarray}
we only need to analyze the numerator. We have
\begin{eqnarray}
 &&(2a^2|h(t)|^2+|b|^2)^2(\sqrt{a^2|h(t)|^4+|b|^2|h(t)|^2}+\sqrt{1-2a|h(t)|^2+a^2|h(t)|^4+|b|^2|h(t)|^2})^2 \nonumber  \\
 &&-4a^2(a^2|h(t)|^4+|b|^2|h(t)|^2)  \nonumber  \\
&&=(4a^4|h(t)|^4+|b|^4+a^2|b|^2|h(t)|^2)(2a^2|h(t)|^4+2|b|^2|h(t)|^2-2a|h(t)|^2 \nonumber \\
&&+2\sqrt{a^2|h(t)|^4+|b|^2|h(t)|^2}\sqrt{1-2a|h(t)|^2+a^2|h(t)|^4+|b|^2|h(t)|^2})+|b|^4. \nonumber
\end{eqnarray}
Because $(4a^4|h(t)|^4+|b|^4+a^2|b|^2|h(t)|^2)>0$ and $|b|^4>0$, we only need to consider
\begin{eqnarray}
 &&2a^2|h(t)|^4+2|b|^2|h(t)|^2-2a|h(t)|^2+2\sqrt{a^2|h(t)|^4+|b|^2|h(t)|^2}\sqrt{1-2a|h(t)|^2+a^2|h(t)|^4+|b|^2|h(t)|^2} \nonumber  \\
&&=2a|h(t)|^2(a|h(t)|^2-1)+2|b|^2|h(t)|^2+2\sqrt{a^2|h(t)|^4+|b|^2|h(t)|^2}\sqrt{(1-a|h(t)|^2)^2+|b|^2|h(t)|^2} \nonumber \\
&&\geq 2a|h(t)|^2(a|h(t)|^2-1)+2|b|^2|h(t)|^2+2\sqrt{a^2|h(t)|^4+|b|^2|h(t)|^2}(1-a|h(t)|^2) \\
&&=(2\sqrt{a^2|h(t)|^4+|b|^2|h(t)|^2}-2a|h(t)|^2)(1-a|h(t)|^2)+2|b|^2|h(t)|^2, \nonumber
\end{eqnarray}
where the equality holding if and only if  $|b|^2|h(t)|^2=0$ for Eq. (35), and the initial state is an incoherent one. Thus we have $2a^2|h(t)|^4+2|b|^2|h(t)|^2-2a|h(t)|^2+2\sqrt{a^2|h(t)|^4+|b|^2|h(t)|^2}\sqrt{1-2a|h(t)|^2+a^2|h(t)|^4+|b|^2|h(t)|^2}>0$, which implies that $G(t)>0$.

\end{document}